\documentclass[
reprint,onecolumn,
superscriptaddress, 12pt, 
nofootinbib,
 amsmath,amssymb,
floatfix
]{revtex4-2}

\usepackage{physics} 
\usepackage{graphicx}
\usepackage{dcolumn}
\usepackage{bm}
\usepackage{cancel}
\usepackage{amsfonts}
\usepackage{xcolor}
\usepackage{tcolorbox}
\usepackage{placeins}
\usepackage{float}

\newcommand{\beqa}{\begin{eqnarray}}
\newcommand{\eeqa}{\end{eqnarray}}
\newcommand{\nn}{\nonumber}
\usepackage{dcolumn}
\usepackage{bm}

\usepackage{amsmath, fullpage,setspace}
\usepackage{graphicx}
\usepackage{subcaption}

\begin{document}

\title{
Collective Dynamics of Vortex Clusters in Compact Fluid Domains:
From Pair Interactions to a Quadrupole Description
}
\author{Aswathy K R}
\affiliation{Birla Institute of Technology and Science, Pilani, Hyderabad Campus, Telangana 500078, India}
\author{Rickmoy Samanta}
\affiliation{Birla Institute of Technology and Science, Pilani, Hyderabad Campus, Telangana 500078, India}
\affiliation{Indian Institute of Technology Kharagpur, West Bengal 721302, India}

\begin{abstract}
Clusters of co-rotating vortices on compact fluid domains exhibit a simple collective dynamics, combining coherent global rotation with a slow  breathing of the cluster size. In this work, we investigate an analytic model of vortex interactions on a doubly periodic inviscid fluid domain, based on an exact representation in terms of the Schottky-Klein prime function and its $q$-representation. The two-vortex problem reduces to a single complex degree of freedom, from which explicit expressions for the orbital rotation frequency and dipole translation velocity are obtained. Building on this framework, we derive a small-cluster expansion that reveals a  decomposition of the dynamics into universal planar interactions, isotropic torus corrections, and geometry-induced anisotropic modes. At leading order, the collective dynamics is encoded in a single complex quadrupole moment: its real part governs corrections to the rotation rate, while its imaginary part controls the slow breathing of the cluster. These predictions are quantitatively confirmed by direct numerical simulations, establishing a reduced description of vortex clusters on the flat torus and compact fluid domains.
\end{abstract}

\maketitle
\section{Introduction}

The study of point vortices in two-dimensional incompressible and inviscid flows has long served as a bridge between discrete vortex models and continuum fluid mechanics. Point vortices provide an idealized description in which vorticity is concentrated at discrete points, reducing complex fluid motion to a finite-dimensional dynamical system. Interest in vortex dipoles, clusters, and related collective structures has grown substantially in recent years, driven by experimental and theoretical developments in quantum fluids, active matter, and hydrodynamic rotor systems. In particular, vortex dipoles and clusters have been observed and characterized in trapped Bose-Einstein condensates and two-dimensional quantum fluids~\cite{Neely2010,Freilich2010,white2012,White2014,vsc}, while related collective phenomena have been explored in superfluid turbulence and vortex dynamics~\cite{Rooney2011,Goodman2015,Stagg2016}, anomalous vortex fluids~\cite{abanov}, and in astrophysical settings such as vortex assemblies in the superfluid interior of neutron stars~\cite{Haskell2015}. Similar collective dynamics also arise in soft-matter and active systems, including plane-confined microrotors and their mixtures~\cite{lushi,yeo}, hydrodynamically interacting rotating inclusions and rotor assemblies~\cite{sh1,sh2}, as well as vortex flows on curved membranes~\cite{sam2021}. These developments provide strong motivation for reduced descriptions of interacting vortex ensembles.

Much of the theoretical and computational literature on vortex dipoles and clusters has focused on unbounded or effectively planar geometries, where translational symmetry and the absence of global image interactions considerably simplify the dynamics~\cite{nc2009}. Compact periodic domains, however, introduce qualitatively new features. In such geometries, each vortex interacts with an infinite lattice of its periodic images, and the resulting motion is governed by the Green’s function of the Laplace operator on the flat torus. This global coupling modifies both few-body motion and collective many-body dynamics, making periodic domains a natural setting in which geometry and topology enter explicitly into vortex interactions.
Point-vortex dynamics in doubly periodic domains has since been explored in detail~\cite{WeissMcWilliams1991}, including analytic descriptions in finite geometries using elliptic-function methods~\cite{Kunin1994}, as well as studies of few-vortex configurations and dipoles emphasizing integrability and the effect of periodic boundary conditions~\cite{StremlerAref1999,Stremler2010,TsangKanso2013}.

A major advance in the analysis of vortex motion in multiply connected domains is the use of the Schottky-Klein prime function, which provides a compact representation of the hydrodynamic Green’s function and incorporates periodic image interactions~\cite{Crowdy2005,Crowdy2016,grms}. This approach yields explicit interaction kernels and  equations of motion for $N$ vortices in doubly periodic geometries~\cite{KrishnamurthySakajo2023}, with an equivalent formulation in terms of the $q$-digamma function $\psi_{\rho}(z)$ enabling efficient computation and extensions such as harmonic fields~\cite{sam3,sam4}.

In parallel, vortex dynamics on compact surfaces has been studied from geometric and topological perspectives, where curvature and global constraints play a central role~\cite{BoattoKoiller2008}. Recent formulations based on harmonic one-forms and Hodge decomposition provide a general framework for flows on higher-genus surfaces~\cite{GrottaRagazzoGustafssonKoiller2024}, while studies of vortex pairs and dipoles reveal connections to magnetic geodesics~\cite{Gustafsson2022,DrivasGlukhovskiyKhesin2024,sam2021,sam2025,sam2026}. For the flat torus, the harmonic component reduces to a constant background flow and can be removed without loss of generality~\cite{GrottaRagazzoGustafssonKoiller2024}, making it especially amenable to analytic treatment.
Motivated by these developments, the present work studies vortex clusters in doubly periodic planar domains using the analytic framework of the Schottky-Klein prime function and its $q$-digamma representation~\cite{sam3}. While periodic geometries introduce additional complexity through global topology and image interactions, they also admit simple analytic structure. Building on recent developments in this direction~\cite{sam3}, we show that this framework extends naturally to large vortex ensembles on the flat torus. At the same time, recent experimental studies of vortex clustering in quantum fluids~\cite{vsc} provide additional motivation for theoretical work in such directions.\\\\
Before proceeding, we note that the present work complements and builds on a substantial body of prior studies on vortex interactions in doubly periodic domains~\cite{WeissMcWilliams1991,Kunin1994,StremlerAref1999,Stremler2010,grms,TsangKanso2013,lushi,KrishnamurthySakajo2023}. It advances this line of research by developing $q$-function-based closed-form expressions for the interaction kernel and associated dynamical quantities, enabling analytic reductions and a coarse-grained description of large-$N$ vortex clusters in compact geometries. In particular, the two-vortex problem is reduced to a single complex degree of freedom, yielding explicit expressions for the orbital rotation frequency and dipole translation velocity, while for many-vortex configurations a decomposition of the dynamics emerges into universal planar interactions, isotropic torus corrections, and geometry-induced anisotropic modes. At leading order, the collective dynamics is governed by a single complex quadrupole moment: its real part controls corrections to the rotation rate, while its imaginary part determines the slow breathing of the cluster.\\\\
The paper is organized as follows. In Sec.~\ref{hm}, we formulate the Hamiltonian structure of point-vortex motion on the flat torus and derive the exact equations of motion in annulus variables, together with the closed-form kernel representation. In Sec.~\ref{fund}, we establish the conserved quantities and use the kernel antisymmetry to obtain an exact reduction of the two-vortex problem, leading to a complete description of binary dynamics. In Sec.~\ref{sec:local_cluster}, we develop a  small-cluster expansion based on the local form of the kernel, yielding a reduced description for compact same-sign vortex clusters. In Sec.~\ref{omegatheory}, we derive a coarse-grained expression for the collective angular velocity in terms of the cluster size and quadrupole moment and detailed numerical validation of these predictions. In Sec.~\ref{sizeevolution}, we obtain an analytic evolution law for the cluster size, showing that its slow variation is governed by the imaginary part of the quadrupole moment. We conclude in Sec.~\ref{cncl}, and relegate technical details on the Schottky-Klein formalism and kernel expansions to the Appendices.

\section{Model setup}
\label{hm}
Motivated by the structure outlined above, we revisit the formulation of the $N$-vortex interactions in a doubly periodic rectangular domain recently introduced by Sakajo and Krishnamurthy~\cite{KrishnamurthySakajo2023}, based on the hydrodynamic Green’s function expressed in terms of the Schottky-Klein prime function. A key advance of~\cite{KrishnamurthySakajo2023} is the derivation of a closed Hamiltonian system valid for arbitrary total circulation, in which the effects of periodicity and background vorticity arise intrinsically from the compact geometry. 

Building on this framework, we exploit its equivalent $q$-digamma representation developed in~\cite{sam3} to recast the dynamics in a form that makes the analytic structure of the interaction kernel explicit, thereby enabling  reductions and coarse-grained descriptions. 

We consider $N$ point vortices in a doubly periodic rectangular domain with periods $2\pi$ and $-\log\rho$ ($0<\rho<1$), with vortex strengths $\Gamma_j$ and complex positions $w_j = x_j + i y_j$. It is well known that the canonical symplectic structure of the point vortex system gives the Hamiltonian equations
\begin{equation}
\Gamma_j \dot{x}_j = \frac{\partial H}{\partial y_j},
\qquad
\Gamma_j \dot{y}_j = -\,\frac{\partial H}{\partial x_j},
\label{eq:real-hamilton}
\end{equation}
where $H$ is the interaction Hamiltonian. Introducing the complex coordinate
$w_j = x_j + i y_j$, we combine \eqref{eq:real-hamilton} to obtain the complex Hamiltonian form
\begin{equation}
\Gamma_j \frac{d w_j}{dt}
= -\,2 i\,\frac{\partial H}{\partial \overline{w}_j}.
\label{eq:complex-hamilton}
\end{equation}
Following the approach of~\cite{grms,KrishnamurthySakajo2023} and related works, we introduce annulus coordinates
\begin{equation}
\nu_j = e^{i w_j}, \qquad
w_j = -\,i \log \nu_j.
\label{eq:w-to-nu}
\end{equation}
and \eqref{eq:complex-hamilton} yields
\begin{equation}
\Gamma_j \frac{d\,\overline{w}_j}{dt}
= -\,2\,\nu_j\,\frac{\partial H}{\partial \nu_j}.
\label{eq:wbar-evolution}
\end{equation}
The vortex Hamiltonian on the flat torus can be written as the sum of pairwise
interactions and a Robin self term:
\begin{equation}
H(\nu_1,\ldots,\nu_N)
=
-\sum_{1\le j<k\le N}\Gamma_j\Gamma_k\,
G\!\left(\frac{\nu_j}{\nu_k};\sqrt{\rho}\right)
-\frac12\sum_{j=1}^N \Gamma_j^2\,\widehat G(\nu_j;\sqrt{\rho}),
\label{eq:H-full}
\end{equation}
with the pairwise Green function  expressed through the Schottky-Klein prime function $P$ defined in terms of a generic variable $\zeta$ (see Appendix~\ref{appsk} for more details on the Schottky-Klein machinery)
\begin{equation}
G(\zeta;\sqrt{\rho})
=
\frac{1}{2\pi}\log\!\bigl|P(\zeta,\sqrt{\rho})\bigr|
-\frac{1}{4\pi}\log|\zeta|
+\frac{1}{4\pi\log\rho}\,\bigl(\log|\zeta|\bigr)^2,
\label{eq:G-prime}
\end{equation}
and the Robin (self) term
\begin{equation}
\widehat G(\nu;\sqrt{\rho})
=
\frac{1}{2\pi}\log\!\Bigg|\prod_{m=1}^{\infty}(1-\rho^{m})^{2}\Bigg|.
\label{eq:Robin}
\end{equation}
For the flat torus the Robin function \eqref{eq:Robin} is independent of \(\nu\) (a
geometry–dependent constant), hence
\begin{equation}
\nu_j\,\frac{\partial}{\partial \nu_j}\widehat G(\nu_j;\sqrt{\rho})=0,
\qquad
\overline{\nu}_j\,\frac{\partial}{\partial \overline{\nu}_j}\widehat G(\nu_j;\sqrt{\rho})=0.
\label{eq:Robin-vanish}
\end{equation}
Therefore only the pairwise part of \eqref{eq:H-full} contributes. Using this fact and inserting \eqref{eq:H-full} and \eqref{eq:G-prime} into the  Hamilton's equation
\eqref{eq:wbar-evolution}
and dividing by \(\Gamma_j\), yields  the fundamental $N$-vortex dynamical equation   describing motion of $N$ point vortices of circulations $\Gamma_1,\dots,\Gamma_N$ in a doubly-periodic rectangular domain with periods $2\pi$ and $-\log\rho$
($0<\rho<1$) (note our original flat coordinates $w_j = x_j + i y_j$ ) as
\begin{equation}
\frac{d\overline{w}_j}{dt}
=
\frac{1}{2\pi}
\sum_{\substack{k=1\\k\neq j}}^{N}
\Gamma_k\,K\!\left(\frac{\nu_j}{\nu_k},\sqrt{\rho}\right)
-\frac{1}{4\pi}
\sum_{\substack{k=1\\k\neq j}}^{N}\Gamma_k
+\frac{1}{2\pi\log\rho}
\sum_{\substack{k=1\\k\neq j}}^{N}
\Gamma_k
\log\!\left|\frac{\nu_j}{\nu_k}\right|,
\label{dyneq}
\end{equation}
where $\nu_j = e^{i w_j}$ are coordinates on the concentric annulus. A closed analytic form for the function $K$, introduced in \cite{sam3}, is given by
\begin{equation}
K(\zeta, \sqrt{\rho})
  = \frac{1}{1 - \zeta}
    + \frac{1}{\log \rho}
      \left[
        \psi_{\rho}\!\left(\frac{\log(1/\zeta)}{\log \rho}\right)
        - \psi_{\rho}\!\left(\frac{\log(\zeta)}{\log \rho}\right)
      \right],
\label{eq:K_qdigamma}
\end{equation}
where $\psi_{\rho}(z)$ denotes the \emph{$q$-digamma function} with base
$q=\rho$, defined by the logarithmic derivative
of the \emph{$q$-gamma function} $\Gamma_{q}(z)$,
\begin{equation}
\psi_{\rho}(z)
  = \frac{d}{dz}\log \Gamma_{\rho}(z),
\qquad
\Gamma_{\rho}(z)
  = (1-\rho)^{1-z}
    \prod_{n=0}^{\infty}
    \frac{1-\rho^{\,n+1}}{1-\rho^{\,n+z}},
\label{eq:qgamma_def}
\end{equation}
which converges for $0<\rho<1$ and all complex $z$. Noting that $\nu_k = e^{i w_k}$, the logarithmic term in~\eqref{dyneq} reduces to
\begin{equation}
\log\left|\frac{\nu_j}{\nu_k}\right| = -(y_j - y_k).\nn
\end{equation}

\section{Fundamental motion of the vortex binary}
\label{fund}
\begin{figure}[t]
\centering

\begin{subfigure}{0.48\textwidth}
\includegraphics[width=\linewidth]{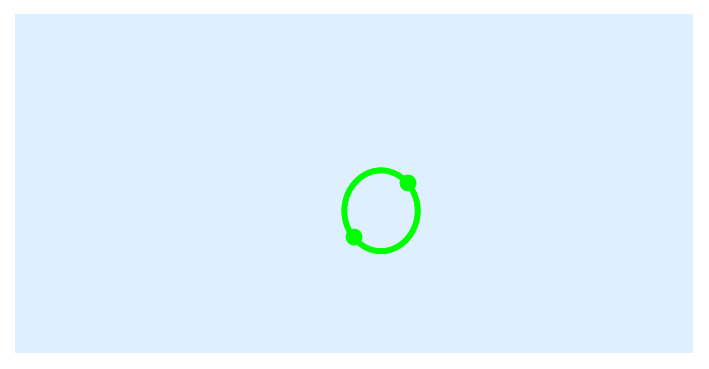}
\caption{Trajectories ($\Gamma_1=\Gamma_2=1$).}
\end{subfigure}
\hfill
\begin{subfigure}{0.48\textwidth}
\includegraphics[width=\linewidth]{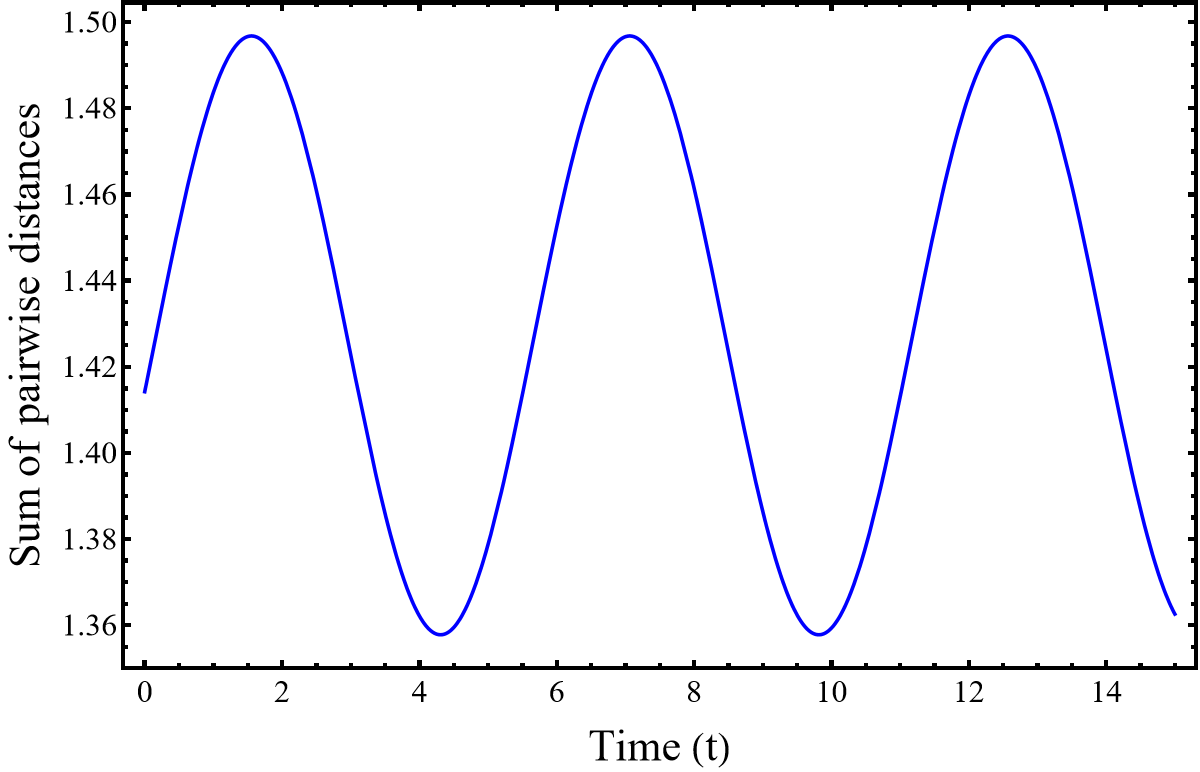}
\caption{Distance (oscillatory).}
\end{subfigure}

\vspace{0.2cm}

\begin{subfigure}{0.48\textwidth}
\includegraphics[width=\linewidth]{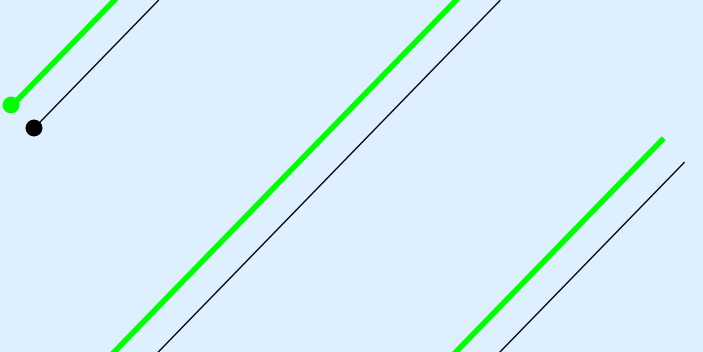}
\caption{Dipole trajectories ($\Gamma_1=-\Gamma_2$).}
\end{subfigure}
\hfill
\begin{subfigure}{0.48\textwidth}
\includegraphics[width=\linewidth]{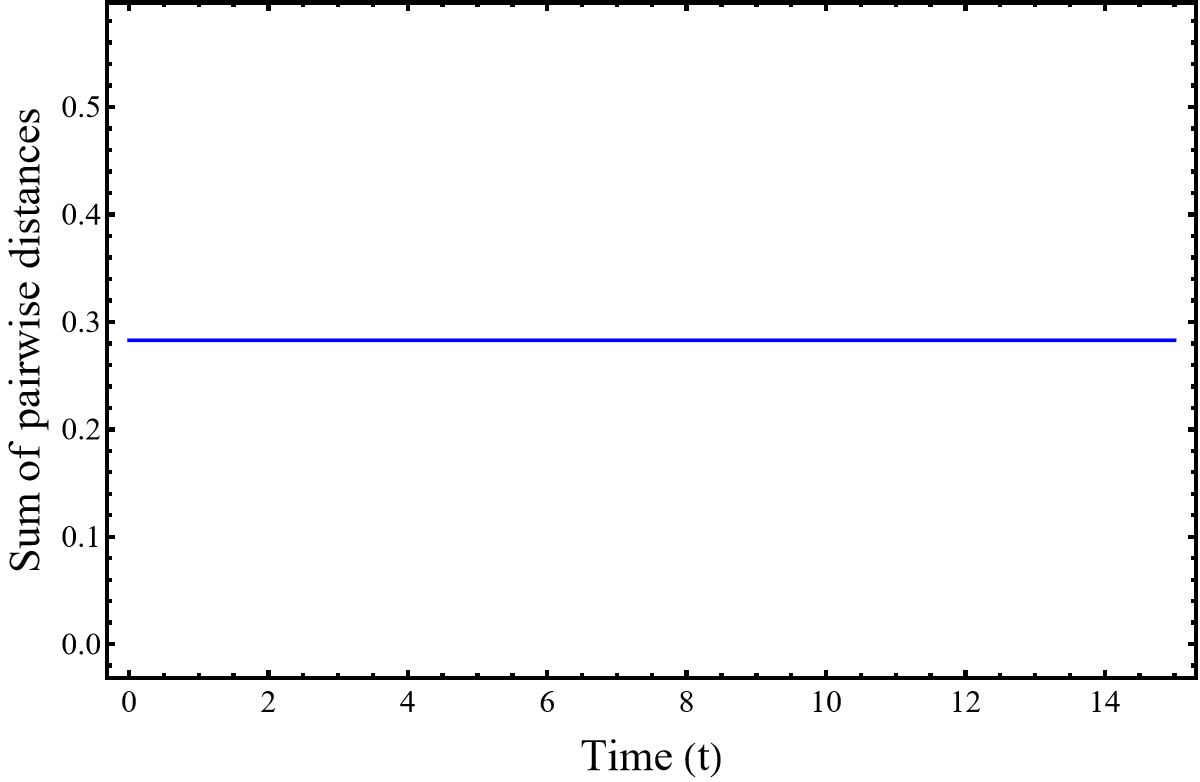}
\caption{Distance (constant).}
\end{subfigure}
\caption{Two-vortex motion on the flat torus. Top row: nonzero total circulation (chiral vortex pair), showing periodic trajectories and oscillatory inter-vortex distance. Bottom row: dipole case ($\Gamma_1=-\Gamma_2$), exhibiting rigid motion with constant separation. Green and black dots indicate the initial positions of vortices with circulations $+1$ and $-1$, respectively; the same color scheme is used for the corresponding trajectories.}
\label{fig:torus_two_vortex_and_dipole}
\end{figure}

\begin{figure}[t]
\centering
\includegraphics[width=0.5\linewidth]{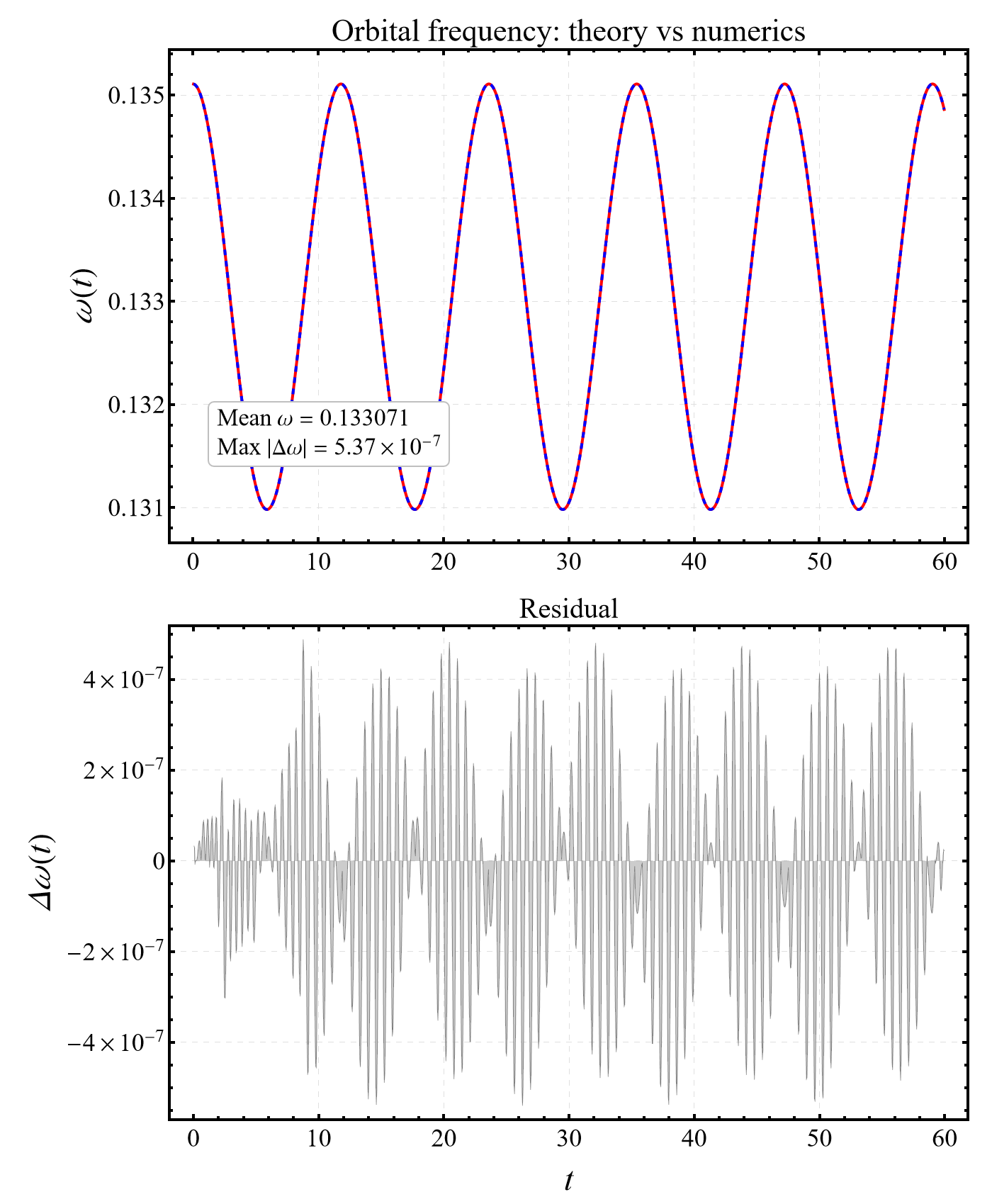}
\caption{Theoretical vs numerical orbital frequency on the flat torus.
Top: $\omega(t)$ (theory and numerics indistinguishable).
Bottom: residual $\Delta\omega \sim 10^{-7}$.}
\label{2vomega}
\end{figure}

\begin{figure}[h!]
\begin{tabular}{lcccccccc}
\includegraphics[width=0.5\linewidth]{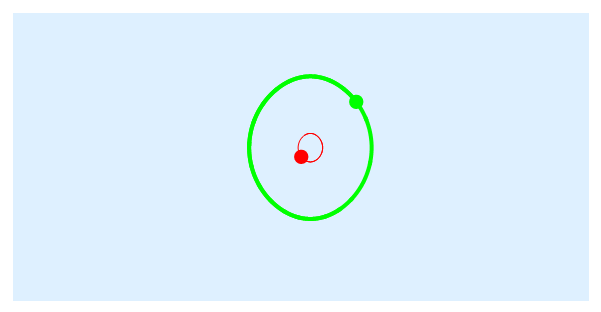}&&\includegraphics[width=0.5\linewidth]{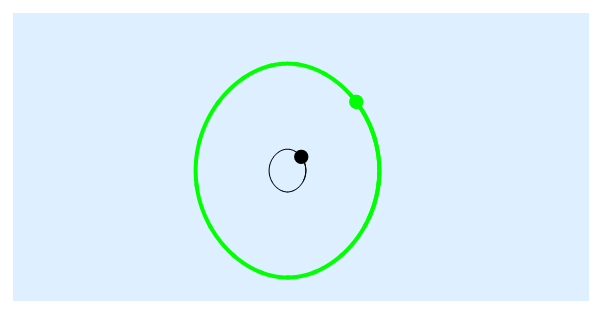}
\end{tabular}
\caption{Motion of vortices of unequal strengths. Green, red, and black dots indicate the initial positions of vortices with circulations $+1$, $+5$, and $-5$, respectively; the same color scheme is used for the corresponding trajectories.}
 \label{uneq} 
\end{figure}
The dissipationless dynamics \eqref{dyneq} inherits the Hamiltonian structure of the flat torus and therefore admits the conserved Hamiltonian \(H\) together with the circulation-weighted centroid
\begin{equation}
C:=\sum_{j=1}^N \Gamma_j w_j,
\qquad
\overline{C}:=\sum_{j=1}^N \Gamma_j \overline{w}_j .
\end{equation}
The conservation of \(H\) follows directly from \eqref{eq:complex-hamilton} and its complex conjugate, while the conservation of \(C\) is obtained by writing \eqref{dyneq} in the form
\begin{equation}
\frac{d\overline{w}_j}{dt}
=
\sum_{\substack{k=1\\k\neq j}}^{N}
\Gamma_k\,F\!\left(\frac{\nu_j}{\nu_k}\right),
\qquad
F(\zeta)
=
\frac{1}{2\pi}K(\zeta,\sqrt{\rho})
-\frac{1}{4\pi}
+\frac{1}{2\pi\log\rho}\log|\zeta|.
\end{equation}
Using $K(1/\zeta)=1-K(\zeta)$ and $\log|1/\zeta|=-\log|\zeta|$ we obtain the antisymmetric relation
\begin{equation}
F(1/\zeta)=-F(\zeta),
\label{eq:F-antisym}
\end{equation}
so that the double sum in \(d\overline{C}/dt\) cancels pairwise. In real variables, the centroid conservation is equivalent to the two translational invariants
\begin{equation}
\sum_{j=1}^{N}\Gamma_j x_j,
\qquad
\sum_{j=1}^{N}\Gamma_j y_j.
\end{equation}
The total circulation is also trivially conserved. Unlike the planar problem, no further invariant associated with continuous rotational symmetry exists on the flat torus.

The same antisymmetry governs relative motion. Defining \(w_{jk}=w_j-w_k\), we have
\begin{equation}
\frac{d\overline{w}_{jk}}{dt}
=
(\Gamma_j+\Gamma_k)\,F\!\left(\frac{\nu_j}{\nu_k}\right)
+
\sum_{\substack{\ell=1\\ \ell\neq j,k}}^{N}
\Gamma_\ell
\left[
F\!\left(\frac{\nu_j}{\nu_\ell}\right)
-
F\!\left(\frac{\nu_k}{\nu_\ell}\right)
\right].
\end{equation}
For \(N=2\) this closes exactly:
\begin{equation}
\frac{d\overline{w}_{12}}{dt}
=
(\Gamma_1+\Gamma_2)\,F\!\left(\frac{\nu_1}{\nu_2}\right).
\label{eq:w12-binary}
\end{equation}
We see immediately from \eqref{eq:w12-binary} that when \(\Gamma_1+\Gamma_2=0\) the relative coordinate is frozen,
\begin{equation}
\frac{d w_{12}}{dt}=0,
\end{equation}
so the two vortices translate rigidly as a dipole with constant separation, as shown in the lower panels of Fig.~\ref{fig:torus_two_vortex_and_dipole}.
For generic vortex strengths, the binary dynamics depends on the total circulation and reduces to a single complex degree of freedom. Introducing the variable $\eta$ defined as
\begin{equation}
\eta:=\frac{\nu_1}{\nu_2}=e^{i(w_1-w_2)},
\label{eq:binary-F}
\end{equation}
For \(\Gamma_{\rm tot}\neq0\), the conserved centroid determines the center of motion and the binary is reconstructed from the relative coordinate \(\Delta:=w_1-w_2\). Since \(\eta=e^{i\Delta}\), \eqref{eq:w12-binary} gives
\begin{equation}
\dot{\eta}
=
i\,\eta\,\Gamma_{\rm tot}\,F(\eta),
\label{eq:eta-binary}
\end{equation}
and hence the quadrature
\begin{equation}
\int_{\eta_0}^{\eta(t)}\frac{d\zeta}{i\,\zeta\,F(\zeta)}
=
\Gamma_{\rm tot}(t-t_0).
\label{eq:binary-quadrature}
\end{equation}
 For \(\Gamma_{\rm tot}\neq0\), the pair  undergoes nontrivial relative motion and the inter-vortex distance oscillates, as illustrated in the upper panels of Fig.~\ref{fig:torus_two_vortex_and_dipole}. The two-vortex problem on the flat torus is therefore completely integrable, with stationary separations determined by
\begin{equation}
F(\eta_*)=0.
\end{equation}
Writing $\eta$ in polar form $\eta=r\,e^{i\theta},
$ Eq.\eqref{eq:eta-binary} yields
\begin{equation}
\frac{\dot r}{r}
=
-\Gamma_{\rm tot}\,\Im F(\eta),
\qquad
\dot\theta
=
\Gamma_{\rm tot}\,\Re F(\eta).
\label{eq:rtheta-binary}
\end{equation}
With
\begin{equation}
K(\eta,\sqrt{\rho})=K_R(\eta,\sqrt{\rho})+i\,K_I(\eta,\sqrt{\rho}),
\end{equation}
this becomes
\begin{align}
\frac{\dot r}{r}
&=
-\frac{\Gamma_{\rm tot}}{2\pi}\,K_I(\eta,\sqrt{\rho}),
\label{eq:r-final-binary}
\\
\dot\theta
&=
\Gamma_{\rm tot}
\left[
\frac{1}{2\pi}K_R(\eta,\sqrt{\rho})
-\frac{1}{4\pi}
+\frac{\log r}{2\pi\log\rho}
\right].
\label{eq:theta-final-binary}
\end{align}
Thus the imaginary part of \(K\) drives radial drift in the annulus, while the real part determines the phase evolution. Constant-\(r\) motions satisfy \(K_I(\eta,\sqrt{\rho})=0\), in which case the binary rotates with angular velocity (in the $\eta$ variable)
\begin{equation}
\Omega_\eta
=
\Gamma_{\rm tot}
\left[
\frac{1}{2\pi}K_R(\eta,\sqrt{\rho})
-\frac{1}{4\pi}
+\frac{\log r}{2\pi\log\rho}
\right].
\label{eq:Omega-eta}
\end{equation}
For equal like-signed vortices, \(\Gamma_1=\Gamma_2=\gamma\), these reduce to
\begin{align}
\frac{\dot r}{r}
&=
-\frac{\gamma}{\pi}\,K_I(\eta,\sqrt{\rho}),
\\
\Omega_\eta
&=
\frac{\gamma}{\pi}
\left[
K_R(\eta,\sqrt{\rho})-\frac12+\frac{\log r}{\log\rho}
\right].
\end{align}
Although \(\theta\) gives the natural annulus phase, the physically observed orbital frequency is more naturally defined in Euclidean coordinates. Writing
\begin{equation}
\Delta=w_1-w_2=x_{12}+iy_{12},
\qquad
\theta_E=\arg(x_{12}+iy_{12}),
\end{equation}
the Euclidean orbital frequency is
\begin{equation}
\Omega_E:=\dot\theta_E
=
\frac{x_{12}\dot y_{12}-y_{12}\dot x_{12}}{x_{12}^2+y_{12}^2}.
\label{eq:OmegaE-def-compact}
\end{equation}
From \eqref{eq:binary-F},
\begin{equation}
\dot x_{12}=\Gamma_{\rm tot}F_R(\eta),
\qquad
\dot y_{12}=-\Gamma_{\rm tot}F_I(\eta),
\end{equation}
and therefore
\begin{equation}
\Omega_E
=
-\Gamma_{\rm tot}
\frac{x_{12}F_I(\eta)+y_{12}F_R(\eta)}{x_{12}^2+y_{12}^2}.\nn
\end{equation}
Using
\begin{equation}
F_R=
\frac{1}{2\pi}K_R(\eta,\sqrt{\rho})
-\frac{1}{4\pi}
-\frac{y_{12}}{2\pi\log\rho},
\qquad
F_I=
\frac{1}{2\pi}K_I(\eta,\sqrt{\rho}),
\end{equation}
one obtains the exact Euclidean frequency formula
\begin{equation}
\Omega_E
=
-\frac{\Gamma_{\rm tot}}{2\pi}
\frac{
x_{12}K_I(\eta,\sqrt{\rho})
+
y_{12}K_R(\eta,\sqrt{\rho})
-\tfrac12 y_{12}
-\dfrac{y_{12}^2}{\log\rho}
}{x_{12}^2+y_{12}^2}.
\label{eq:OmegaE-final-compact}
\end{equation}
Equation~\eqref{eq:OmegaE-final-compact} shows that the Euclidean orbital frequency is distinct from the annulus-phase frequency \(\dot\theta\). For the representative equal-vortex orbit shown in Fig.~\ref{2vomega}, the theoretical prediction from \eqref{eq:OmegaE-final-compact} is visually indistinguishable from the numerically extracted frequency, with residuals at the \(10^{-7}\) level. The numerical decomposition reveals that the Euclidean orbital frequency
is controlled by three contributions (orbit averaged),
\begin{equation}
\Omega_E \propto
\left\langle
\frac{x_{12}K_{\mathrm I}
+
y_{12}K_{\mathrm R}
-\tfrac12 y_{12}-
y_{12}^2/\log\rho
}{x_{12}^2+y_{12}^2}
\right\rangle.
\end{equation}
While the term proportional to $y_{12}K_{\mathrm R}$ provides the dominant
contribution, the correction arising from $x_{12}K_{\mathrm I}$ is
numerically significant, and the geometric term proportional to
$y_{12}^2/\log\rho$ yields a smaller but non-negligible contribution.
This demonstrates that the orbital frequency is not determined solely
by the mean separation, but arises from correlated oscillations between
the relative geometry and the interaction kernel. For the trajectories considered here, the fluctuation of the Euclidean
radius is sufficiently weak that
\begin{equation}
\left\langle \frac{A}{x_{12}^2+y_{12}^2} \right\rangle
\approx
\frac{\langle A\rangle}{\langle x_{12}^2+y_{12}^2\rangle},
\end{equation}
where
\begin{equation}
A :=
x_{12}K_{\mathrm I}(\eta,\sqrt{\rho})
+
y_{12}K_{\mathrm R}(\eta,\sqrt{\rho})
-\frac{1}{2}y_{12}
-\frac{y_{12}^2}{\log\rho}.
\end{equation}
This yields the compact approximation
\begin{equation}
\Omega_E
\approx
-\frac{\Gamma_{\rm tot} \langle A\rangle}{2\pi\,\langle x_{12}^2+y_{12}^2\rangle}
\label{eq:Omega_compact_mean}
\end{equation}
In the dipole case,
\begin{equation}
\Gamma_1=\gamma,
\qquad
\Gamma_2=-\gamma,
\qquad
\Gamma_{\rm tot}=0,\nn
\end{equation}
the relative coordinate is constant and both vortices move with the same velocity,
\begin{equation}
\frac{d\overline{w}_1}{dt}
=
\frac{d\overline{w}_2}{dt}
=
-\gamma\,F(\eta).
\end{equation}
Writing \(d\overline{w}/dt=\dot x-i\dot y\), one obtains
\begin{align}
\dot x
&=
-\gamma
\left[
\frac{1}{2\pi}K_R(\eta,\sqrt{\rho})
-\frac{1}{4\pi}
+\frac{\log|\eta|}{2\pi\log\rho}
\right],
\\
\dot y
&=
\frac{\gamma}{2\pi}K_I(\eta,\sqrt{\rho}).
\end{align}
Since \(\eta=e^{i(x_{12}+iy_{12})}\), we have \(\log|\eta|=-y_{12}\), and the dipole velocity may be written as
\begin{align}
\dot x
&=
-\frac{\gamma}{4\pi}
\left[
2K_R(\eta,\sqrt{\rho})
-1
-\frac{2y_{12}}{\log\rho}
\right],
\\
\dot y
&=
\frac{\gamma}{2\pi}K_I(\eta,\sqrt{\rho}).
\label{eq:dipole-velocity-final}
\end{align}
Thus the dipole speed on the flat torus is not determined solely by the separation magnitude, as in the planar problem, but also by the periodic image effects encoded in \(K_R\) and \(K_I\), together with the explicit geometric correction proportional to \(y_{12}/\log\rho\). This sensitivity to global geometry is reflected in the rigid dipole trajectories shown in Fig.~\ref{fig:torus_two_vortex_and_dipole}. More generally, binaries with unequal strengths can be treated in exactly the same way: the dynamics again reduces to the single complex variable $\eta=\nu_1/\nu_2$. Representative trajectories for unequal like-signed and opposite-signed pairs are shown in Fig.~\ref{uneq}.

\section{Same sign cluster dynamics}
\label{sec:local_cluster}
Clusters of like-signed vortices on the flat torus exhibit a simple but nontrivial collective behavior. To leading order, the cluster undergoes a coherent rotation driven by the mutual pairwise interactions, while its overall size remains nearly constant. Superimposed on this rotation is a slower modulation of the cluster size, corresponding to a weak breathing mode induced by the compact geometry. These two effects—rotation and slow breathing—capture the dominant large-scale dynamics of tightly bound vortex clusters.

To quantify this collective behavior systematically, we derive a small-cluster expansion of the vortex motion on the flat torus by expanding the closed analytic form of the interaction kernel
\begin{equation}
K(\zeta,\sqrt{\rho})
=
\frac{1}{1-\zeta}
+
\frac{1}{\log \rho}
\left[
\psi_{\rho}\!\left(\frac{\log(1/\zeta)}{\log \rho}\right)
-
\psi_{\rho}\!\left(\frac{\log(\zeta)}{\log \rho}\right)
\right],
\end{equation}
where $\psi_{\rho}$ is the $q$-digamma function. Introducing multiplicative and additive variables as before (repeated here for convenience)
\begin{equation}
\zeta=e^{z}, \qquad z=i w, \qquad w=x+i y,\nn
\end{equation}
it is convenient to expand in $z=\log\zeta$ about coincidence ($z\to0$). This yields
\begin{equation}
K(z)
=
\frac{1}{z}
+
\frac{1}{2}
+
\left(
\frac{1}{12}
-
\frac{2\,\psi_{\rho}^{(1)}(1)}{\log^2 \rho}
\right) z
+
O(z^3),
\end{equation}
where $\psi_{\rho}^{(1)}(1)=QPolyGamma(1,1;\rho)$ (see Appendix~\ref{app:locexp} for a detailed and rather subtle derivation). Passing to the physical coordinate $z=i w$, the kernel becomes
\begin{equation}
K(w)
=
-\frac{i}{w}
+
\frac{1}{2}
+
\left(
\frac{i}{12}
-
\frac{2 i\,\psi_{\rho}^{(1)}(1)}{\log^2 \rho}
\right) w
+
O(w^3).
\end{equation}
The interaction kernel entering the equations of motion is
\begin{equation}
F(w,\bar w)
=
\frac{1}{2\pi} K(w)
-
\frac{1}{4\pi}
+
\frac{i}{4\pi \log\rho}(w-\bar w),\nn
\end{equation}
which is equivalent to the  form
\begin{equation}
F(\zeta)
=
\frac{1}{2\pi}K(\zeta,\sqrt{\rho})
-
\frac{1}{4\pi}
+
\frac{1}{2\pi\log\rho}\log|\zeta|.
\end{equation}
Substituting the local expansion gives
\begin{equation}
F(w,\bar w)
=
-\frac{i}{2\pi w}
+
A(\rho)\,w
+
B(\rho)\,\bar w
+
\cdots,
\end{equation}
with
\begin{equation}
A(\rho)
=
\frac{i}{24\pi \log^2 \rho}
\left[
\log\rho(6+\log\rho)
-
24\,\psi_{\rho}^{(1)}(1)
\right],
\qquad
B(\rho)
=
-\frac{i}{4\pi \log\rho}.
\end{equation}
The leading term recovers the planar interaction,
\begin{equation}
F(w,\bar w)\sim -\frac{i}{2\pi w}, \qquad w\to0,
\end{equation}
while the coefficients $A(\rho)$ and $B(\rho)$ encode the torus geometry. We now consider a compact cluster of $N$ vortices of equal circulation $\Gamma$ and introduce internal coordinates
\begin{equation}
w_j = R + \xi_j, \qquad \sum_{j=1}^N \xi_j = 0.\nn
\end{equation}
Using the conservation law $\dot R=0$  and substitution into the equations of motion yields
\beqa
\dot{\bar \xi}_j
=
\sum_{k\neq j}
\Gamma
\left[
-\frac{i}{2\pi(\xi_j-\xi_k)}
+
A(\rho)(\xi_j-\xi_k)
+
B(\rho)(\bar\xi_j-\bar\xi_k)
\right].\nn\\
=
-\frac{i\Gamma}{2\pi}
\sum_{k\neq j}\frac{1}{\xi_j-\xi_k}
+
N\Gamma A(\rho)\,\xi_j
+
N\Gamma B(\rho)\,\bar\xi_j.
\eeqa
This expression shows that the dynamics of a compact same-sign vortex cluster decomposes into a universal planar interaction together with geometry-induced corrections determined by the torus modulus $\rho$. The $B(\rho)$ term contributes an isotropic component to the collective motion, while the $A(\rho)$ term generates an anisotropic deformation that couples directly to the cluster shape.
\section{Coarse-grained angular velocity of a clustered configuration}
\label{omegatheory}
\begin{figure*}[t]
\centering
\includegraphics[width=0.12\textwidth]{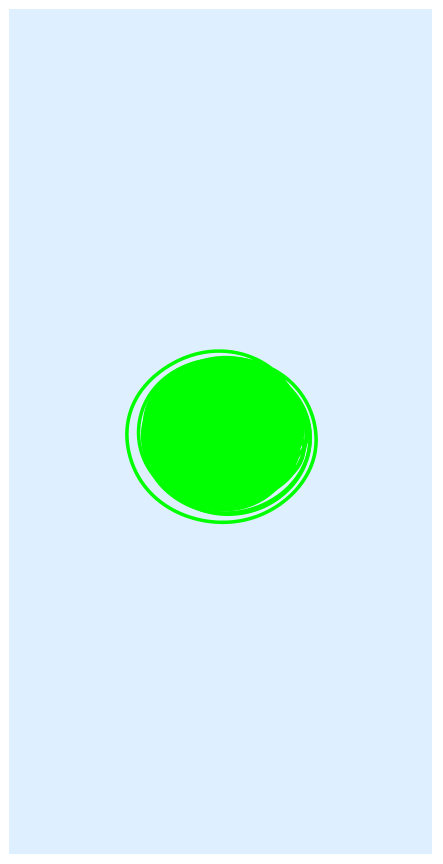}
\hspace{0.01\textwidth}
\includegraphics[width=0.40\textwidth]{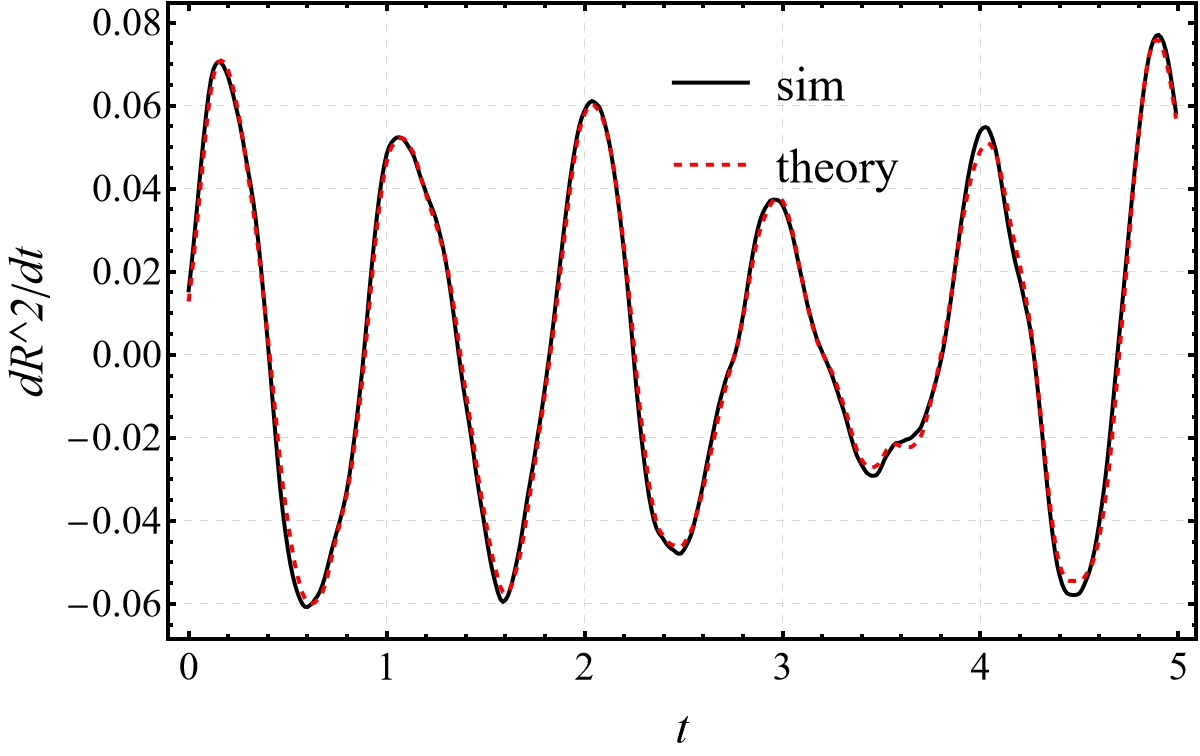}
\hspace{0.01\textwidth}
\includegraphics[width=0.40\textwidth]{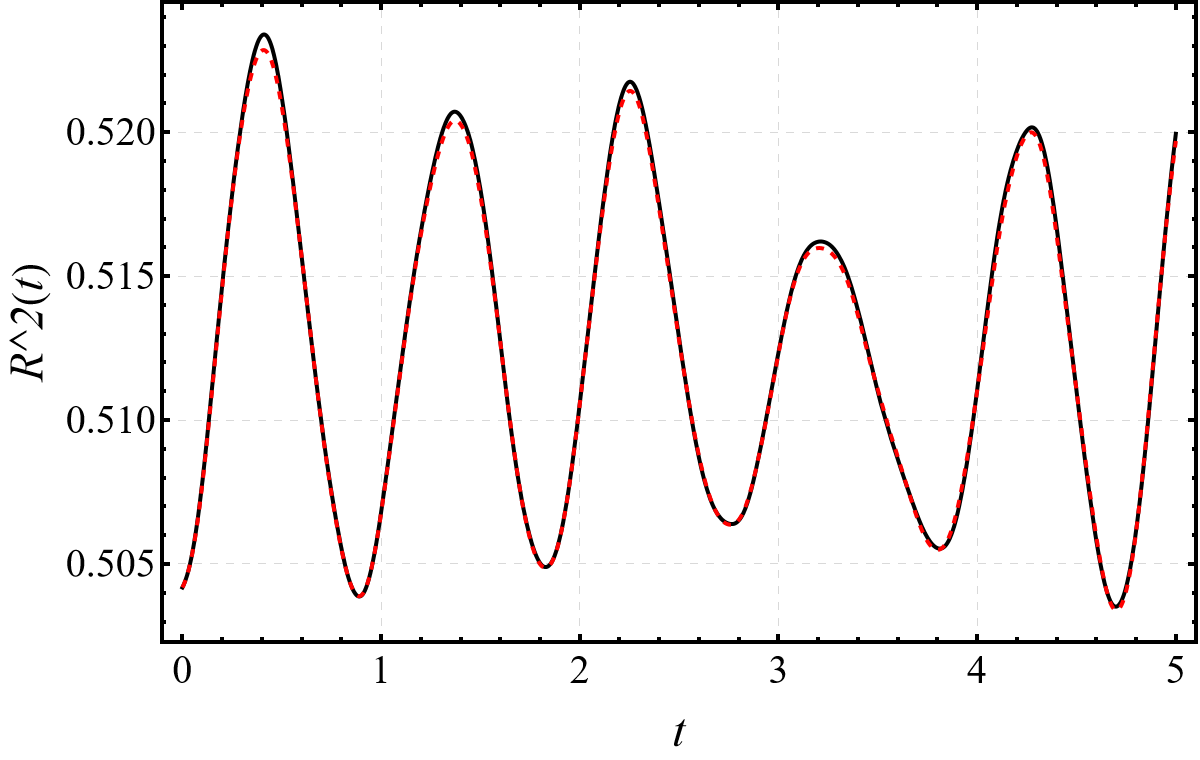}
\caption{Numerical  test of size-evolution for a compact vortex cluster on the flat torus ($\rho=e^{-4\pi}$; $N=50$ equal-circulation vortices randomly initialized within a disk in the fundamental cell). From left to right: real-space trajectories showing that the configuration remains compact;  comparison of the simulated $dR^2/dt$ with the theoretical prediction $-2\Gamma A_I(\rho)\,\mathrm{Im}\,Q(t)$; and comparison of the simulated $R^2(t)$ with the integrated theoretical prediction. The close agreement in the last two panels confirms that the weak breathing of the cluster is accurately governed by the imaginary part of the quadrupole moment. The Hamiltonian variation satisfies $\Delta H \lesssim 10^{-4}$ over the duration of the simulation.}
\label{fig:size_diagnostics}
\end{figure*}
\begin{figure}[t]
\centering
\includegraphics[width=0.40\linewidth]{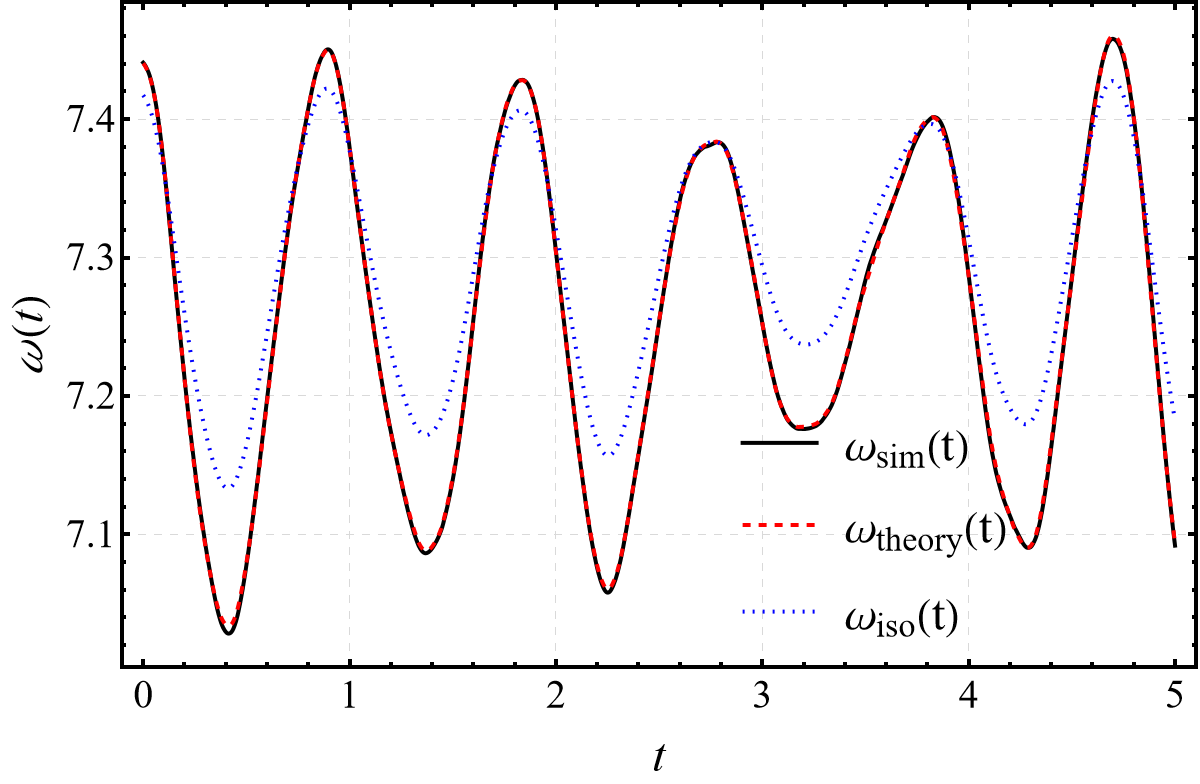}
\includegraphics[width=0.40\linewidth]{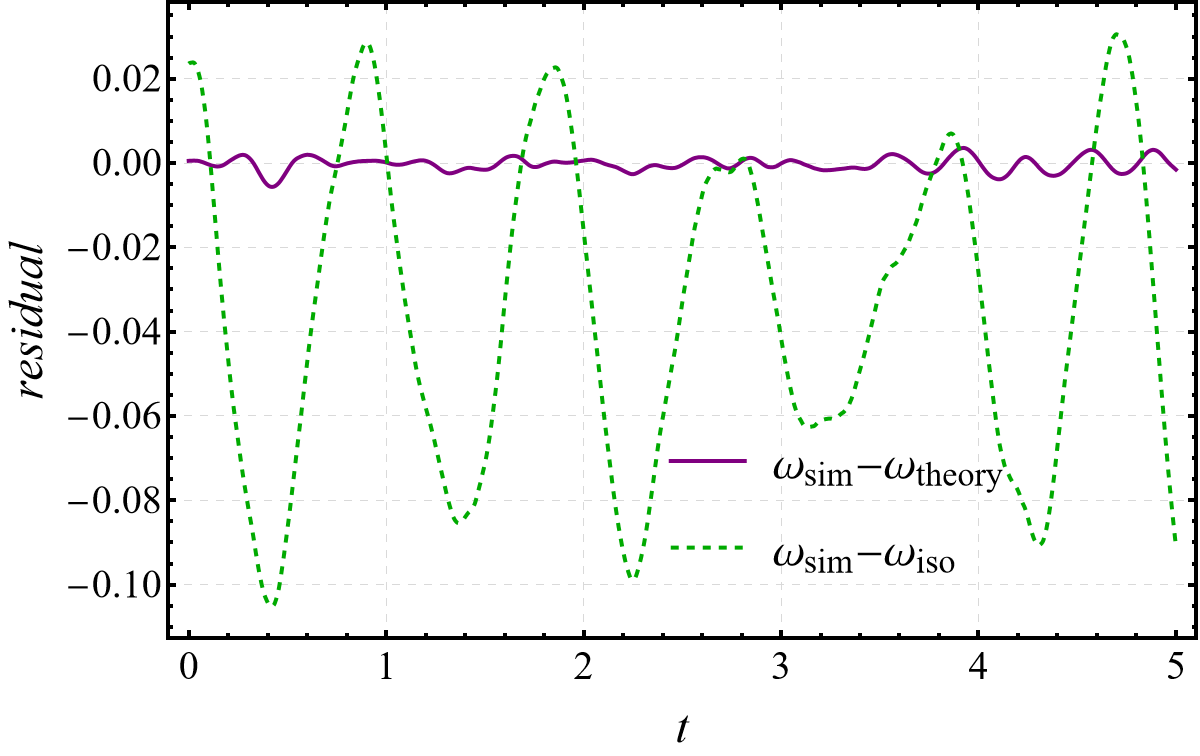}

\medskip

\includegraphics[width=0.40\linewidth]{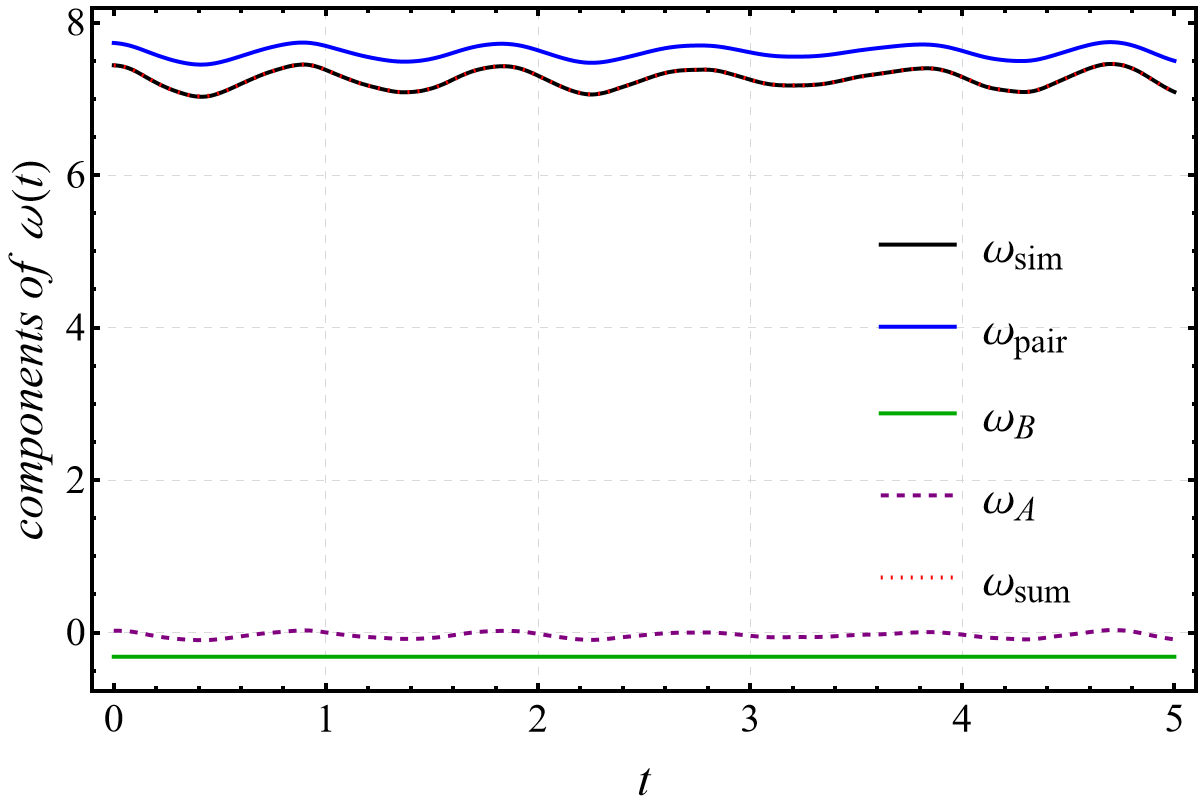}
\includegraphics[width=0.40\linewidth]{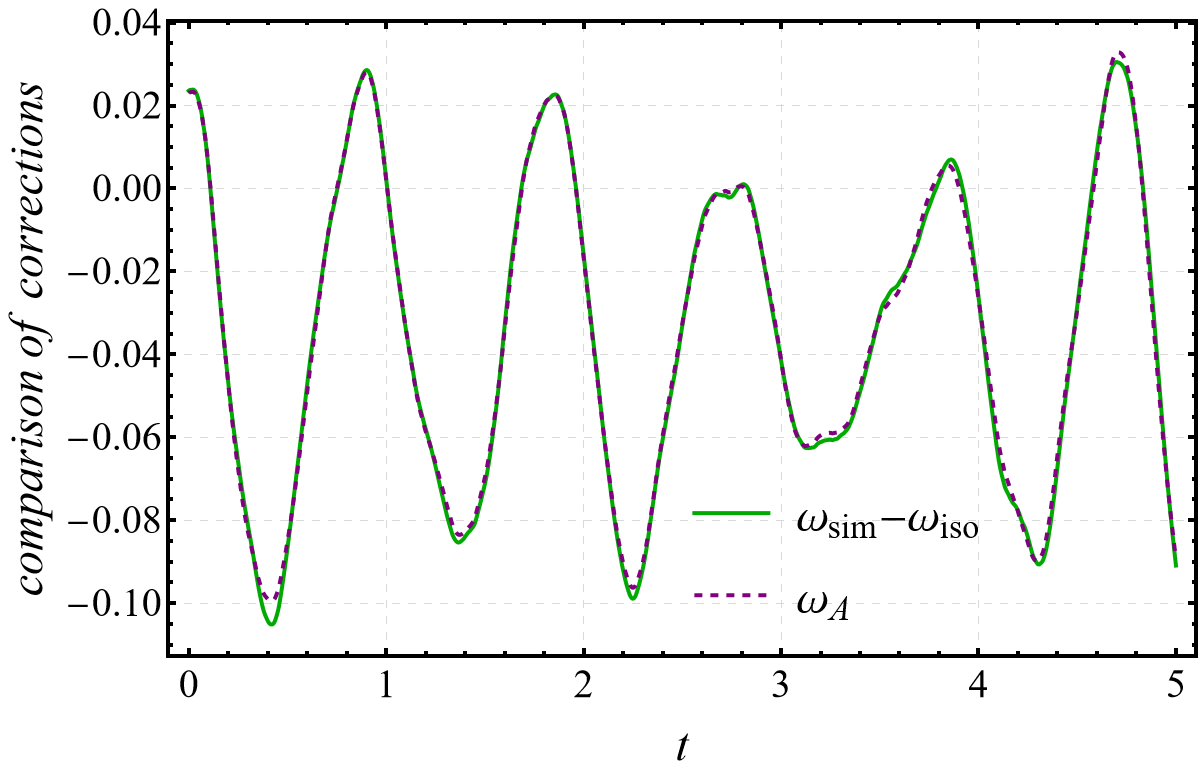}

\medskip

\includegraphics[width=0.40\linewidth]{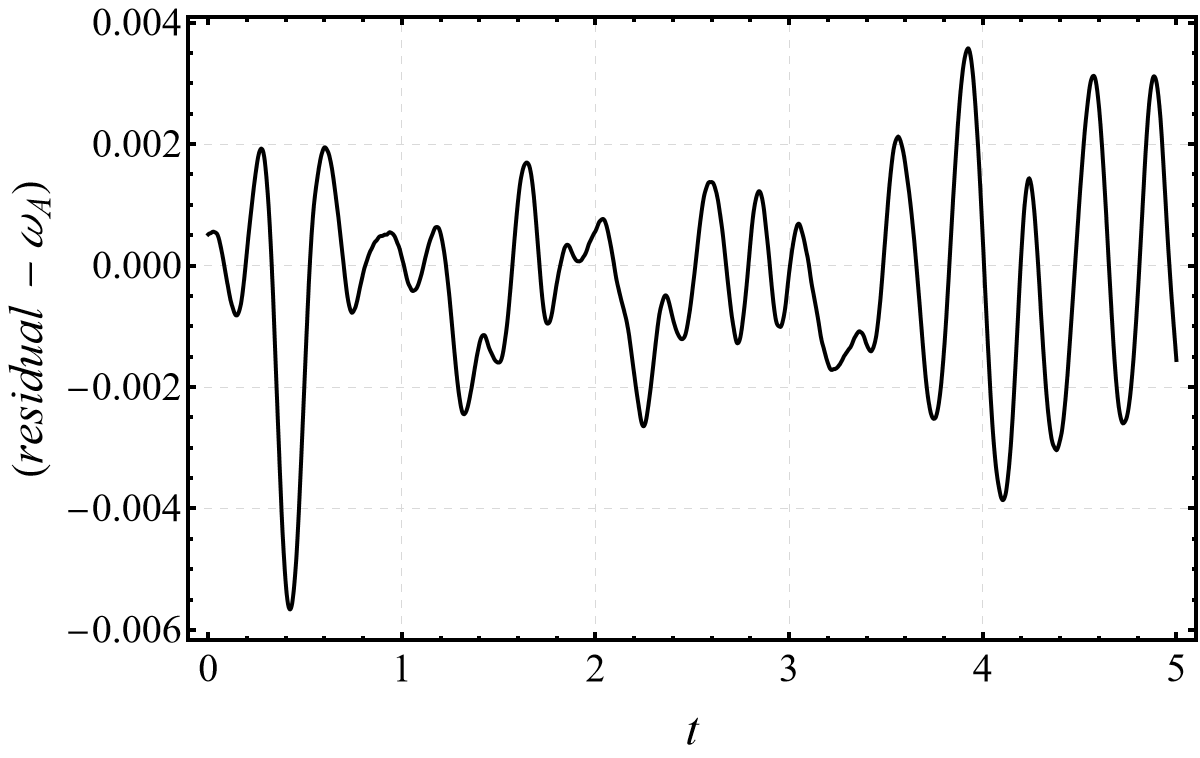}

\caption{Comparison between numerical simulations and coarse-grained theory for a compact vortex cluster on the flat torus ($\rho=e^{-4\pi}$; $N=50$ equal-circulation vortices randomly initialized in a disk). 
Top left: $\Omega_{\rm sim}(t)$ vs $\Omega_{\rm theory}(t)$ and $\Omega_{\rm iso}(t)$. 
Top right: residuals. 
Middle left: decomposition into pair, torus, and quadrupolar contributions. 
Middle right: $\Omega_{\rm sim}-\Omega_{\rm iso}$ vs $\Omega_A(t)$. 
Bottom: difference, showing that the quadrupolar term captures the deviation from isotropic theory.}
\label{fig:omega_comparison}
\end{figure}
We can now extract a collective angular velocity of the vortex cluster directly from the reduced dynamics without assuming any cluster symmetry. Defining
\begin{equation}
I:=\sum_{j=1}^N |\xi_j|^2,
\qquad
Q:=\sum_{j=1}^N \xi_j^2,
\end{equation}
we now proceed to calculate the rotation rate
\begin{equation}
\Omega(t)
=
\frac{\sum_{j=1}^N \operatorname{Im}(\bar\xi_j \dot\xi_j)}{I}
=
-\frac{\operatorname{Im}\!\left(\sum_{j=1}^N \xi_j \dot{\bar\xi}_j\right)}{I}.
\end{equation}
Using the reduced dynamics of the previous section (presented here for clarity),
\begin{equation}
\dot{\bar \xi}_j
=
-\frac{i\Gamma}{2\pi}
\sum_{k\neq j}\frac{1}{\xi_j-\xi_k}
+
N\Gamma A(\rho)\,\xi_j
+
N\Gamma B(\rho)\,\bar\xi_j,
\end{equation}
we obtain
\begin{equation}
\sum_{j=1}^N \xi_j \dot{\bar\xi}_j
=
-\frac{i\Gamma}{2\pi}
\sum_{j=1}^N\sum_{k\neq j}\frac{\xi_j}{\xi_j-\xi_k}
+
N\Gamma A(\rho)\,Q
+
N\Gamma B(\rho)\,I.
\end{equation}
The double sum evaluates to $\tfrac12 N(N-1)$, yielding
\begin{equation}
\Omega(t)
=
\frac{\Gamma N(N-1)}{4\pi I}
-
\frac{\operatorname{Im}\!\bigl(N\Gamma A(\rho)\,Q\bigr)}{I}
-
\frac{\operatorname{Im}\!\bigl(N\Gamma B(\rho)\,I\bigr)}{I}.
\end{equation}
Writing $R^2=I/N$, $B(\rho)=-i/(4\pi\log\rho)$, and $A(\rho)=iA_I(\rho)$, this reduces to
\begin{equation}
\Omega(t)
=
\frac{\Gamma (N-1)}{4\pi R^2}
+
\frac{N\Gamma}{4\pi\log\rho}
-
\frac{N\Gamma A_I(\rho)\operatorname{Re}(Q)}{I},
\end{equation}
where
\begin{equation}
A_I(\rho)
=
\frac{1}{24\pi \log^2 \rho}
\left[
\log\rho(6+\log\rho)
-
24\,\psi_{\rho}^{(1)}(1)
\right].
\end{equation}
Thus the collective angular velocity decomposes into a universal planar term, an isotropic torus-induced shift, and an anisotropic correction governed by the quadrupole moment. In the nearly isotropic cluster  $Q\approx 0$, one obtains
\begin{equation}
\Omega(t)
\approx
\frac{\Gamma (N-1)}{4\pi R^2}
+
\frac{N\Gamma}{4\pi\log\rho}.
\end{equation}
The leading term in the coarse-grained angular velocity plays the role of a many-body analogue of the rigid two-rotor frequency in an unbounded 3D fluid (see Appendix~\ref{app:rotors} and Ref.~\cite{lushi} for a typical soft matter setup), with the cluster radius replacing the pair separation and the effective strength set by $\Gamma(N-1)$. Notably, the scaling differs: $\Omega\sim R^{-2}$ in the planar vortex system, as opposed to $\Omega\sim D^{-3}$ for 3D rotors. The additional torus-induced shift and quadrupolar correction are specific to the periodic geometry and to the internal deformability of the cluster.
We compare the coarse-grained prediction for the angular velocity with
direct numerical simulations of the full vortex motion on the flat torus.
The simulations are initialized with $N=50$ equal-circulation vortices
randomly distributed within a compact disk in the fundamental cell of a
rectangular torus with $\rho=e^{-4\pi}$, ensuring a tightly clustered,
anisotropic configuration with $Q(0)\neq 0$. The Hamiltonian variation satisfies $\Delta H \lesssim 10^{-4}$ over the duration of the simulation. The theoretically derived formula
\begin{equation}
\Omega_{\rm cg}(t)
=
\frac{\Gamma (N-1)}{4\pi R^2(t)}
+
\frac{N\Gamma}{4\pi\log\rho}
-
\frac{N\Gamma A_I(\rho)\,\operatorname{Re}(Q(t))}{I(t)},
\end{equation}
with $I(t)=\sum_j |\xi_j|^2$, $R^2(t)=I(t)/N$, and $Q(t)=\sum_j \xi_j^2$ can be compared with a numerically extracted solution
\begin{equation}
\Omega_{\rm sim}(t)
=
\frac{\sum_{j=1}^N \operatorname{Im}(\bar\xi_j \dot\xi_j)}{\sum_{j=1}^N |\xi_j|^2},
\end{equation}
and compare it with the full $\Omega_{\rm cg}(t)$ and the isotropic approximation
\begin{equation}
\Omega_{\rm iso}(t)
=
\frac{\Gamma (N-1)}{4\pi R^2(t)}
+
\frac{N\Gamma}{4\pi\log\rho}.
\end{equation}
As shown in Fig.~\ref{fig:omega_comparison}, $\Omega_{\rm cg}(t)$ is in
near-perfect agreement with $\Omega_{\rm sim}(t)$ over the full evolution,
while $\Omega_{\rm iso}(t)$ exhibits a clear deviation. The
residual $\Omega_{\rm sim}-\Omega_{\rm cg}$ remains at the level of
$10^{-3}$, whereas the deviation from the isotropic prediction is an order
of magnitude larger, demonstrating that the quadrupolar correction is
essential for quantitative accuracy. A physically motivated way to decompose the rotation rate is
\begin{equation}
\Omega(t)
=
\Omega_{\rm pair}(t)
+
\Omega_B
+
\Omega_A(t),
\end{equation}
with
\begin{equation}
\Omega_{\rm pair}(t)=\frac{\Gamma (N-1)}{4\pi R^2(t)},
\qquad
\Omega_B=\frac{N\Gamma}{4\pi\log\rho},
\qquad
\Omega_A(t)=-\frac{N\Gamma A_I(\rho)\operatorname{Re}(Q(t))}{I(t)}.
\end{equation}
In Fig.~\ref{fig:omega_comparison}, we find that $\Omega_{\rm pair}$ provides the dominant contribution, while
$\Omega_B$ produces a constant geometric offset. The remaining
time-dependent modulation is entirely captured by $\Omega_A(t)$, which
accurately reproduces the difference $\Omega_{\rm sim}-\Omega_{\rm iso}$.
The residual
\begin{equation}
(\Omega_{\rm sim}-\Omega_{\rm iso})-\Omega_A
\end{equation}
remains small throughout the evolution, confirming that higher-order
corrections are negligible in the compact-cluster dynamics.

The size evolution is independently validated in
Fig.~\ref{fig:size_diagnostics}, where the numerical results for
$dR^2/dt$ and $R^2(t)$ agree closely with the analytic prediction
$-2\Gamma A_I(\rho)\operatorname{Im}(Q)$ (derived in next section). These results demonstrate that
the leading-order theory accurately captures both the collective rotation
and the weak breathing of the cluster, with the quadrupole moment $Q(t)$
providing the key dynamical control of deviations from isotropic motion.
\section{Analytic evolution of the cluster size}
\label{sizeevolution}
The reduced dynamics of the previous section also implies that the cluster size  is tied to the quadrupole moment. Defining
\begin{equation}
I(t):=\sum_{j=1}^N |\xi_j|^2,
\qquad
R^2(t)=\frac{I(t)}{N},\nn
\end{equation}
we differentiate to obtain
\begin{equation}
\dot I
=
2\,\operatorname{Re}\!\left(\sum_{j=1}^N \xi_j \dot{\bar\xi}_j\right).\nn
\end{equation}
Using the reduced equation of motion,
\begin{equation}
\dot{\bar \xi}_j
=
-\frac{i\Gamma}{2\pi}
\sum_{k\neq j}\frac{1}{\xi_j-\xi_k}
+
N\Gamma A(\rho)\,\xi_j
+
N\Gamma B(\rho)\,\bar\xi_j,\nn
\end{equation}
we find
\begin{equation}
\sum_{j=1}^N \xi_j \dot{\bar\xi}_j
=
-\frac{i\Gamma}{2\pi}
\sum_{j=1}^N\sum_{k\neq j}\frac{\xi_j}{\xi_j-\xi_k}
+
N\Gamma A(\rho)\,Q
+
N\Gamma B(\rho)\,I,
\end{equation}
where \(Q=\sum_{j=1}^N \xi_j^2\). The double sum evaluates to \(\tfrac12 N(N-1)\), so that the pair contribution is purely imaginary and does not affect \(\dot I\). The term proportional to \(B(\rho)=-i/(4\pi\log\rho)\) is also purely imaginary. The only contribution to \(\dot I\) therefore arises from the term involving \(A(\rho)=iA_I(\rho)\), giving
\begin{equation}
\dot I
=
-2N\Gamma A_I(\rho)\,\operatorname{Im}(Q).
\end{equation}
It follows that
\begin{equation}
\frac{dR^2}{dt}
=
-2\Gamma A_I(\rho)\,\operatorname{Im}(Q).
\label{eq:R2dot_final}
\end{equation}
Thus the cluster size evolves solely through the imaginary part of the quadrupole moment. In particular, when \(\operatorname{Im}(Q)\) is small or oscillatory with small mean, the cluster exhibits only weak breathing and remains approximately of constant size. Together with the angular-velocity law, this shows that the same complex quadrupole governs both collective rotation and size modulation, with \(\operatorname{Re}(Q)\) controlling rotation and \(\operatorname{Im}(Q)\) controlling the breathing dynamics.
To quantitatively validate the size evolution law \eqref{eq:R2dot_final}, we compare its predictions with direct numerical simulations of vortex motion on the flat torus. The simulations are performed for a compact cluster of $N=50$ vortices of equal circulation, initialized randomly within a small disk inside the fundamental domain, with modulus $\rho=e^{-4\pi}$. The initial configuration is thus strongly localized, ensuring that the small-cluster expansion underlying the reduced dynamics is applicable throughout the evolution.

\medskip

The results are summarized in Fig.~\ref{fig:size_diagnostics}. The leftmost panel shows the real-space trajectories, confirming that the vortices remain tightly clustered and do not disperse across the torus.  The Hamiltonian variation satisfies $\Delta H \lesssim 10^{-4}$ over the duration of the simulation. The third panel provides a direct comparison between the numerical time derivative $dR^2/dt$ and the theoretical prediction $-2\Gamma A_I(\rho)\,\operatorname{Im}Q(t)$. The agreement is excellent: the two curves coincide in phase, amplitude, and turning points, demonstrating that the instantaneous rate of change of the cluster size is correctly captured by the reduced quadrupole description.

\medskip

An even more stringent test is obtained by integrating \eqref{eq:R2dot_final} in time. The rightmost panel compares the measured $R^2(t)$ with the reconstructed theoretical prediction
\[
R^2_{\rm theory}(t)
=
R^2(t_0)
-
2\Gamma A_I(\rho)\int_{t_0}^{t}\operatorname{Im}Q(s)\,ds,
\]
and again shows near-perfect agreement over the full evolution. This confirms that not only the instantaneous dynamics, but also the cumulative effect of the quadrupolar term is accurately described by the theory.

\medskip

Together, these results show that the weak temporal variation of the cluster size is governed by the imaginary part of the quadrupole moment, consistent with \eqref{eq:R2dot_final}.  Combined with the angular-velocity relation, this indicates that the leading collective dynamics of compact vortex clusters is encoded in the complex quadrupole $Q(t)$: its real part controls deviations from the shape-independent collective rotation rate, while its imaginary part governs the slow breathing of the cluster.
\section{Conclusion}
\label{cncl}

We have investigated an exact Hamiltonian formulation of point-vortex motion on the flat torus, governed by a closed interaction kernel that incorporates both the singular planar interaction and global geometric effects. The antisymmetry property of the kernel under inversion underlies the conservation laws and enables  reductions of the dynamics. In particular, the two-vortex problem is completely integrable, with a clear distinction between rigid dipole motion and nontrivial chiral dynamics.

A local expansion of the kernel yields a reduced description of compact same-sign vortex clusters, in which the dynamics separates into universal planar, isotropic torus, and anisotropic contributions. This leads to a coarse-grained formulation in terms of collective variables, where the evolution is governed by the second moments of the configuration. The leading correction to the rotation rate is controlled by the real part of the quadrupole moment, while the slow evolution of the cluster size is governed by its imaginary part, identifying the complex quadrupole as the key dynamical quantity encoding deviations from isotropic motion.

Numerical simulations confirm these predictions. For compact clusters, the theoretical angular velocity and size evolution laws accurately reproduce both the mean behavior and the subleading corrections, demonstrating that anisotropy provides the dominant deviation from isotropic dynamics.

These results establish a unified framework linking exact Hamiltonian structure, reduced dynamics, and emergent collective behavior for vortices on the flat torus and periodic fluid domains in general. The formalism presented here will be of interest to the growing body of work on vortex and rotor clusters in superfluid and soft matter systems \cite{nc2009,Neely2010,Freilich2010,Rooney2011,Goodman2015,white2012,White2014,Stagg2016,abanov,vsc,lushi,yeo,sh1,sh2,sam2021}. The formulation naturally suggests extensions to dissipative dynamics, interacting clusters, and more general compact geometries, and provides a pathway toward continuum and kinetic descriptions of vortex matter~\cite{abanov,vsc} in compact domains. 
\section{Acknowledgments}
We are very thankful to Takashi Sakajo, Suryateja Gavva, Naomi Oppenheimer and Haim Diamant.
R.S is supported by DST INSPIRE Faculty fellowship, India (Grant No.IFA19-PH231). Both authors acknowledge support from NFSG and OPERA Research Grant from Birla Institute of Technology and Science, Pilani (Hyderabad Campus).
\section{Data Availability}
The data that support the findings of this study are available within the article. The computational codes used in this work are available from the authors upon reasonable request.

\appendix
\section{Recap of the Schottky-Klein machinery}
\label{appsk}
The Schottky-Klein prime function is a special function defined on
multiply-connected circular domains~\cite{Crowdy2005}.
For the annulus
\[
D_\zeta = \{\zeta \in \mathbb{C} \,|\, \rho < |\zeta| < 1 \},
\]
the prime function is given by the infinite product
\begin{equation}
P(\zeta,\sqrt{\rho})
= (1 - \zeta)
\prod_{k=1}^{\infty}
\!\left(1 - \rho^{k}\zeta\right)
\!\left(1 - \rho^{k}/\zeta\right).
\label{eq:Pfunction}
\end{equation}
Note that $P(\zeta,\sqrt{\rho})$ has a simple zero at $\zeta = 1$ in $D_\zeta$.
The associated $K$-function, defined as the logarithmic derivative of
$P(\zeta,\sqrt{\rho})$, is
\begin{equation}
K(\zeta,\sqrt{\rho})
= \frac{\zeta\,P'(\zeta,\sqrt{\rho})}{P(\zeta,\sqrt{\rho})},
\label{eq:Kdef}
\tag{61}
\end{equation}
where the prime denotes differentiation with respect to the first argument,
i.e.\ $P'(\zeta,\sqrt{\rho}) = \dfrac{d P(\zeta,\sqrt{\rho})}{d\zeta}$. A closed analytic form for the above has been provided in \cite{sam3}:
\begin{equation}
K(\zeta, \sqrt{\rho})
  = \frac{1}{1 - \zeta}
    + \frac{1}{\log \rho}
      \left[
        \psi_{\rho}\!\left(\frac{\log(1/\zeta)}{\log \rho}\right)
        - \psi_{\rho}\!\left(\frac{\log(\zeta)}{\log \rho}\right)
      \right],
\label{eq:K_qdigamma}
\end{equation}
where $\psi_{\rho}(z)$ denotes the \emph{$q$-digamma function} with base
$q=\rho$, defined by the logarithmic derivative
of the \emph{$q$-gamma function} $\Gamma_{\rho}(z)$,
\begin{equation}
\psi_{\rho}(z)
  = \frac{d}{dz}\log \Gamma_{\rho}(z),
\qquad
\Gamma_{\rho}(z)
  = (1-\rho)^{1-z}
    \prod_{n=0}^{\infty}
    \frac{1-\rho^{\,n+1}}{1-\rho^{\,n+z}},
\label{eq:qgamma_def}
\end{equation}
which converges for $0<\rho<1$ and all complex $z$ away from its poles.
In this product representation, each factor in the denominator
$(1-\rho^{\,n+z})$ corresponds to the  lattice of poles
of $\Gamma_{\rho}(z)$. The poles of $\Gamma_{\rho}(z)$ arise from the zeros of the denominator term $(1-\rho^{\,n+z})=0$, which yield
\begin{equation}
\rho^{\,n+z}=1 \;\;\Rightarrow\;\; (n+z)\ln\rho = 2\pi i k, \qquad k\in\mathbb{Z}.
\end{equation}
Thus, the poles are located at
\begin{equation}
z_{n,k} = -\,n + \frac{2\pi i k}{\ln\rho}, \qquad n,k\in\mathbb{Z}.
\end{equation}
These poles lie on a rectangular lattice in the complex $z$–plane, with unit spacing along the real direction and vertical spacing $2\pi/|\ln\rho|$. 
\section{Local expansion of the kernel $K(z)$}
\label{app:locexp}
We derive the small-$z$ expansion of the kernel
\begin{equation}
K(z)
=
\frac{1}{1-e^z}
+
\frac{1}{\log\rho}
\left[
\psi_{\rho}\!\left(-\frac{z}{\log\rho}\right)
-
\psi_{\rho}\!\left(\frac{z}{\log\rho}\right)
\right],
\qquad 0<\rho<1.
\end{equation}
Writing $L=\log\rho$ and $x=z/L$, we expand each contribution about $z=0$.

The elementary term gives
\begin{equation}
\frac{1}{1-e^z}
=
-\frac{1}{z}
+\frac{1}{2}
-\frac{z}{12}
+O(z^3).
\end{equation}
For the $q$-digamma part, we use the shift identity
\begin{equation}
\psi_\rho(u+1)-\psi_\rho(u)
=
-\frac{(\log\rho)\,\rho^u}{1-\rho^u},
\end{equation}
which yields
\begin{equation}
\psi_\rho(-x)-\psi_\rho(x)
=
\bigl[\psi_\rho(1-x)-\psi_\rho(1+x)\bigr]
+
L\!\left[
\frac{e^{-Lx}}{1-e^{-Lx}}
-
\frac{e^{Lx}}{1-e^{Lx}}
\right].
\end{equation}
Expanding for small $x$ gives
\begin{equation}
\psi_\rho(1-x)-\psi_\rho(1+x)
=
-2\,\psi_\rho^{(1)}(1)\,x+O(x^3),\nn
\end{equation}
and
\begin{equation}
L\!\left[
\frac{e^{-Lx}}{1-e^{-Lx}}
-
\frac{e^{Lx}}{1-e^{Lx}}
\right]
=
\frac{2}{x}
+
\frac{L^2}{6}\,x
+O(x^3).\nn
\end{equation}
Combining these results and using $x=z/L$, we obtain
\begin{equation}
\frac{1}{L}\bigl[\psi_\rho(-x)-\psi_\rho(x)\bigr]
=
\frac{2}{z}
+
\left(
\frac{1}{6}
-
\frac{2\,\psi_\rho^{(1)}(1)}{\log^2\rho}
\right)z
+O(z^3).
\end{equation}
Adding the two contributions yields
\begin{equation}
K(z)
=
\frac{1}{z}
+
\frac{1}{2}
+
\left(
\frac{1}{12}
-
\frac{2\,\psi_\rho^{(1)}(1)}{\log^2\rho}
\right)z
+O(z^3),
\end{equation}
which is the local expansion quoted in the main text.

\section{Two rotors in an infinite 3D fluid}
\label{app:rotors}
We consider two rotors aligned with $\hat{\mathbf z}$, with strengths $\Gamma_1$ and $\Gamma_2$ and positions $\mathbf X_i=(x_i,y_i,z_i)$. The induced velocity is
\begin{equation}
\mathbf U(\mathbf X,\mathbf X_j)
=
\Gamma_j\,\hat{\mathbf z}\times
\frac{\mathbf X-\mathbf X_j}{\|\mathbf X-\mathbf X_j\|^3},
\end{equation}
leading to
\begin{equation}
\dot{\mathbf X}_1
=
\Gamma_2\,\hat{\mathbf z}\times\frac{\mathbf X_1-\mathbf X_2}{R^3},
\qquad
\dot{\mathbf X}_2
=
\Gamma_1\,\hat{\mathbf z}\times\frac{\mathbf X_2-\mathbf X_1}{R^3},
\end{equation}
with $R=\|\mathbf X_1-\mathbf X_2\|$. The cross-product structure implies $\dot z_1=\dot z_2=0$, so the motion is planar.

Introducing complex coordinates $z_i=x_i+i y_i$ and $w=z_1-z_2$, we obtain
\begin{equation}
\dot w
=
i(\Gamma_1+\Gamma_2)\frac{w}{|w|^3}.
\end{equation}
Hence $|w|=D$ is constant, so the pair is rigid, and
\begin{equation}
\frac{d}{dt}(\Gamma_1 z_1+\Gamma_2 z_2)=0.
\end{equation}
For $\Gamma_1+\Gamma_2\neq 0$, the conserved center of vorticity
\begin{equation}
C=\frac{\Gamma_1 z_1+\Gamma_2 z_2}{\Gamma_1+\Gamma_2}
\end{equation}
is fixed, and the solution is
\begin{equation}
w(t)=D\,e^{i\Omega t},
\qquad
\Omega=\frac{\Gamma_1+\Gamma_2}{D^3}.
\end{equation}
Writing
\begin{equation}
z_1=C+R_1 e^{i\Omega t},\qquad
z_2=C-R_2 e^{i\Omega t},
\end{equation}
the radii satisfy
\begin{equation}
R_1=\frac{\Gamma_2}{\Gamma_1+\Gamma_2}D,
\qquad
R_2=\frac{\Gamma_1}{\Gamma_1+\Gamma_2}D.
\end{equation}
Thus unequal strengths lead to circular motion about $C$ with unequal radii.

For $\Gamma_1=\Gamma_2=\Gamma$, one recovers symmetric rotation with $R_1=R_2=D/2$ and $\Omega=2\Gamma/D^3$. For $\Gamma_1+\Gamma_2=0$, one finds $\dot w=0$ and
\begin{equation}
\dot z_1=\dot z_2=-\,i\,\Gamma\,\frac{w_0}{D^3},
\end{equation}
so the pair translates rigidly without rotation.

\section{Infinite-plane limit of the orbital frequency}
In this appendix we verify that the exact expression for the Euclidean orbital angular frequency of a vortex binary on the flat torus reduces, in the infinite-plane limit, to the classical result for point vortices in an unbounded domain. We begin from the exact formula derived in the main text for the orbital angular frequency in Euclidean coordinates,
\begin{equation}
\Omega_E =
-\frac{\Gamma_{\mathrm{tot}}}{2\pi}
\frac{
x_{12} K_I(\eta,\sqrt{\rho})
+ y_{12} K_R(\eta,\sqrt{\rho})
-\frac{1}{2}y_{12}
-\frac{y_{12}^2}{\log \rho}
}{
x_{12}^2 + y_{12}^2
},
\label{eq:OmegaE_exact_appendix}
\end{equation}
where $\Delta = w_1 - w_2 = x_{12} + i y_{12}$, $r^2 = x_{12}^2 + y_{12}^2$, and $K = K_R + i K_I$ denotes the interaction kernel. We now consider the infinite-plane limit, corresponding to $\rho \to 0$, so that $\log \rho \to -\infty$. In this limit, the explicit geometric contribution vanishes,
\begin{equation}
\frac{y_{12}^2}{\log \rho} \to 0,
\end{equation}
and the orbital frequency reduces to
\begin{equation}
\Omega_E =
-\frac{\Gamma_{\mathrm{tot}}}{2\pi}
\frac{
x_{12} K_I + y_{12} K_R - \frac{1}{2}y_{12}
}{r^2}.
\label{eq:OmegaE_reduced}
\end{equation}
To evaluate this expression, we use the local expansion of the kernel for small relative separation, as derived in Appendix B of the main text. Writing $w = x_{12} + i y_{12}$, the kernel admits the expansion
\begin{equation}
K(w) = -\frac{i}{w} + \frac{1}{2} + O(w),
\label{eq:K_local_appendix}
\end{equation}
where the higher-order terms encode torus corrections that vanish in the infinite-plane limit. We now compute the real and imaginary parts of the leading-order contribution. Using
\begin{equation}
\frac{1}{w} = \frac{x_{12} - i y_{12}}{r^2},
\end{equation}
we obtain
\begin{equation}
-\frac{i}{w}
=
-\frac{i(x_{12} - i y_{12})}{r^2}
=
\frac{-y_{12} - i x_{12}}{r^2}.
\end{equation}
Thus,
\begin{equation}
K_R = \frac{1}{2} - \frac{y_{12}}{r^2}, 
\qquad
K_I = -\frac{x_{12}}{r^2}.
\end{equation}
Substituting into the numerator of (\ref{eq:OmegaE_reduced}), we find
\begin{align}
x_{12} K_I + y_{12} K_R - \frac{1}{2}y_{12}
&=
x_{12}\left(-\frac{x_{12}}{r^2}\right)
+
y_{12}\left(\frac{1}{2} - \frac{y_{12}}{r^2}\right)
-
\frac{1}{2}y_{12} \nonumber\\
&=
-\frac{x_{12}^2}{r^2}
+
\frac{1}{2}y_{12}
-
\frac{y_{12}^2}{r^2}
-
\frac{1}{2}y_{12}.
\end{align}
The linear terms in $y_{12}$ cancel exactly, yielding
\begin{equation}
x_{12} K_I + y_{12} K_R - \frac{1}{2}y_{12}
=
-\frac{x_{12}^2 + y_{12}^2}{r^2}
=
-1.
\end{equation}

Substituting back into (\ref{eq:OmegaE_reduced}), we obtain
\begin{equation}
\Omega_E =
\left(-\frac{\Gamma_{\mathrm{tot}}}{2\pi}\right)
\left(\frac{-1}{r^2}\right)
=
\frac{\Gamma_{\mathrm{tot}}}{2\pi r^2}.
\end{equation}

To make contact with the classical planar result, we now derive the orbital frequency directly from the Biot–Savart law in the infinite plane. Consider two point vortices of strengths $\Gamma_1$ and $\Gamma_2$ located at positions $\mathbf{r}_1$ and $\mathbf{r}_2$ in an unbounded two–dimensional incompressible fluid. The velocity induced at $\mathbf{r}_1$ by the vortex at $\mathbf{r}_2$ is given by the planar Biot–Savart law,
\begin{equation}
\dot{\mathbf{r}}_1
=
\frac{\Gamma_2}{2\pi}
\frac{\hat{\mathbf{z}} \times (\mathbf{r}_1 - \mathbf{r}_2)}{|\mathbf{r}_1 - \mathbf{r}_2|^2},
\qquad
\dot{\mathbf{r}}_2
=
\frac{\Gamma_1}{2\pi}
\frac{\hat{\mathbf{z}} \times (\mathbf{r}_2 - \mathbf{r}_1)}{|\mathbf{r}_1 - \mathbf{r}_2|^2}.
\end{equation}
Introducing the relative coordinate
\begin{equation}
\mathbf{r} := \mathbf{r}_1 - \mathbf{r}_2,
\qquad r = |\mathbf{r}|,\nn
\end{equation}
we obtain
\begin{align}
\dot{\mathbf{r}}
&=
\dot{\mathbf{r}}_1 - \dot{\mathbf{r}}_2 \nonumber\\
&=
\frac{\Gamma_2}{2\pi}
\frac{\hat{\mathbf{z}} \times \mathbf{r}}{r^2}
-
\frac{\Gamma_1}{2\pi}
\frac{\hat{\mathbf{z}} \times (-\mathbf{r})}{r^2} \nonumber\\
&=
\frac{\Gamma_1 + \Gamma_2}{2\pi}
\frac{\hat{\mathbf{z}} \times \mathbf{r}}{r^2}
=
\frac{\Gamma_{\mathrm{tot}}}{2\pi}
\frac{\hat{\mathbf{z}} \times \mathbf{r}}{r^2}.
\end{align}
This shows that the relative velocity is everywhere perpendicular to $\mathbf{r}$, implying that the separation $r$ is constant in time:
\begin{equation}
\frac{d}{dt} r^2 = 2\,\mathbf{r} \cdot \dot{\mathbf{r}} = 0.
\end{equation}
The motion is therefore a rigid rotation. To extract the angular frequency, we use the kinematic definition
\begin{equation}
\Omega_E = \frac{\mathbf{r} \times \dot{\mathbf{r}}}{r^2}.
\end{equation}
Substituting the expression for $\dot{\mathbf{r}}$, we obtain
\begin{align}
\mathbf{r} \times \dot{\mathbf{r}}
&=
\frac{\Gamma_{\mathrm{tot}}}{2\pi r^2}
\mathbf{r} \times (\hat{\mathbf{z}} \times \mathbf{r}) \nonumber\\
&=
\frac{\Gamma_{\mathrm{tot}}}{2\pi r^2}
\left[
\hat{\mathbf{z}}\,(\mathbf{r}\cdot\mathbf{r})
-
\mathbf{r}\,(\mathbf{r}\cdot\hat{\mathbf{z}})
\right] \nonumber\\
&=
\frac{\Gamma_{\mathrm{tot}}}{2\pi r^2}
\hat{\mathbf{z}}\,r^2,
\end{align}
where we used the vector identity $\mathbf{a}\times(\mathbf{b}\times\mathbf{c})=\mathbf{b}(\mathbf{a}\cdot\mathbf{c})-\mathbf{c}(\mathbf{a}\cdot\mathbf{b})$ and $\mathbf{r}\cdot\hat{\mathbf{z}}=0$. Therefore,
\begin{equation}
\Omega_E
=
\frac{|\mathbf{r} \times \dot{\mathbf{r}}|}{r^2}
=
\frac{\Gamma_{\mathrm{tot}}}{2\pi r^2}.
\end{equation}
This establishes explicitly that the two–vortex system in the infinite plane undergoes uniform circular motion with constant angular frequency proportional to $\Gamma_{\mathrm{tot}}$ and inversely proportional to the square of the separation. In the dipole case $\Gamma_{\mathrm{tot}}=0$, the relative velocity vanishes, and the pair translates rigidly with constant separation, consistent with the standard planar dynamics.
This calculation shows that the exact torus expression (\ref{eq:OmegaE_exact_appendix}) reduces smoothly to the planar result, with the explicit geometric term vanishing and the constant contribution in the kernel canceling the corresponding term in the numerator. The recovery of the classical limit provides a nontrivial consistency check of the formulation.

\end{document}